%% file: oversamp-globe2013.tex
\documentclass[conference]{IEEEtran}
\usepackage{epstopdf}

\usepackage{amsmath}%
\usepackage{amsfonts}%
\usepackage{amssymb}%
\usepackage{graphicx}

\usepackage{pdflscape}
\usepackage{multirow}
\newcommand{\Ts}{T_{\text{symb}}}  
 
\newcommand{\blk}{{n_{\text{symb}}}}		
\newcommand{\nsymb}{{n_{\text{symb}}}}		
\newcommand{\nsamp}{{n}}		

\newcommand{\X}{\textbf{X}}
\newcommand{\Y}{\textbf{Y}}

\newcommand{\x}{\textbf{x}}
\newcommand{\y}{\textbf{y}}

\newcommand{\SNR}{\textsf{SNR}}
\newcommand{\Xsymbol}{{X}_{\text{symb}}}

\newcommand{\Xsymb}[1]{{X}_{\text{symb},#1}}
\newcommand{\xsymb}[1]{{x}_{\text{symb},#1}}

\usepackage{pgf}
\usepackage{tikz}
\usetikzlibrary{arrows,automata}
\usepackage[latin1]{inputenc}
\usepackage{verbatim}

\begin{document}

\title{Multi-sample Receivers Increase Information Rates\\ for Wiener Phase Noise Channels}

\author{
\IEEEauthorblockN{Hassan Ghozlan}
\IEEEauthorblockA{Department of Electrical Engineering\\
University of Southern California\\
Los Angeles, CA 90089 USA\\
ghozlan@usc.edu}
\and
\IEEEauthorblockN{Gerhard Kramer}
\IEEEauthorblockA{
Institute for Communications Engineering \\
Technische Universit\"{a}t M\"{u}nchen \\
80333 Munich, Germany \\
gerhard.kramer@tum.de}
}

\maketitle

\begin{abstract}
A waveform channel is considered where the transmitted signal is corrupted by Wiener phase noise
and additive white Gaussian noise (AWGN).
A discrete-time channel model is introduced that is based on a multi-sample receiver.
Tight lower bounds on the information rates achieved by the multi-sample receiver are computed by means of numerical simulations.
The results show that oversampling at the receiver is beneficial 
for both strong and weak phase noise at high signal-to-noise ratios.
The results are compared with results obtained when using other discrete-time models.
\end{abstract}

\section{Introduction}
Communication systems often suffer from phase noise that arises, e.g., 
due to the instability of RF oscillators 
in satellite \cite{Barbieri2011} or 
microwave links \cite{DurisiICC2013}.
In optical fiber communication, phase noise arises due to the instability of laser oscillators \cite{Foschini1988Comm} 
or due to cross-phase modulation (XPM) in Wavelength-Division-Multiplexing (WDM) systems \cite{JLT2010}.

The nature of the phase noise depends on the application.
A commonly studied \emph{discrete-time} model is
\begin{align}
 Y_k = \Xsymb{k} ~ e^{j \Theta_k} + Z_k
\label{eq:dt-pn-ch}
\end{align}
where $\{Y_k\}$ are the output symbols, $\{\Xsymb{k}\}$ are the input symbols, $\{\Theta_k\}$ is the phase noise process and 
$\{Z_k\}$ is additive white Gaussian noise (AWGN).
For example,
Katz and Shamai \cite{Katz2004} studied the model (\ref{eq:dt-pn-ch}) when $\{\Theta_k\}$ is 
independent and identically distributed (i.i.d.) according to $p_{\Theta}(\cdot)$,
when $\Theta$ is uniformly distributed (called a noncoherent AWGN  channel)
and 
when $\Theta$ has a Tikhonov (or von Mises) distribution (called a partially-coherent AWGN channel).
Tikhonov phase noise models the residual phase error in systems 
with phase-tracking devices, e.g., phase-locked loops (PLL) and ideal interleavers/deinterlevers.

Tight lower bounds on the capacities of memoryless noncoherent and partially coherent AWGN channels 
were computed by solving an optimization problem numerically in \cite{Katz2004} and \cite{Hou2003PCAWGN}, respectively.
Dauwels and Loeliger \cite{Dauwels2008} proposed
a particle filtering method to compute information rates for discrete-time continuous-state channels with memory
and applied the method to (\ref{eq:dt-pn-ch}) for Wiener phase noise and autoregressive--moving-average (ARMA) phase noise.
Barletta, Magarini and Spalvieri \cite{BarlettaLB}
computed lower bounds on information rates for (\ref{eq:dt-pn-ch}) with Wiener phase noise
by using the auxiliary channel technique proposed in \cite{Arnold2006}
and they computed upper bounds in \cite{BarlettaUB}.
They also developed a lower bound based on Kalman filtering in \cite{Barletta2012KalmanLB}.
Barbieri and Colavolpe \cite{Barbieri2011}
computed lower bounds with an auxiliary channel slightly different from \cite{BarlettaLB}.


In this paper, we study a \emph{waveform} channel corrupted  
by Wiener phase noise and AWGN:
\begin{align}
  r(t) = x(t) \ e^{j \theta(t)} + n(t),
  \text{ for } t \in \mathbb{R}
  \label{eq:waveform_ch}
\end{align}
where $x(t)$ and $r(t)$ are the transmitted and received signals, respectively,
while $n(t)$ and $\theta(t)$ are the additive and phase noise, respectively.
A detailed description of the model is given in Sec. \ref{sec:ct-model}.
This model is reasonable, for example, for optical fiber communication with low to intermediate power and laser phase noise,
see \cite{Foschini1988Comm}.
As pointed out in \cite{GhozlanISIT2013}, 
the discrete-time model (\ref{eq:dt-pn-ch}) does not fit the channel (\ref{eq:waveform_ch})
because filtering a phase-varying signal with a constant amplitude gives rise to an output with a varying \emph{amplitude}.
The effect of filtering persists for phase impairments other than Wiener phase noise, e.g., for XPM in optical fiber \cite{GhozlanGVM2011}.
We developed in \cite{GhozlanISIT2013} a discrete-time channel model based on a multi-sample receiver, i.e., 
a filter whose output is sampled multiple times per symbol.

In this paper, we use techniques based on \cite{Arnold2006} to compute 
tight lower bounds on the information rates for the multi-sample receiver introduced in \cite{GhozlanISIT2013}.
The paper is organized as follows.
The continuous-time model is described in Sec. \ref{sec:ct-model}
and the discrete-time model of the multi-sample receiver is described in Sec. \ref{sec:dt-model}.
We develop a method to compute lower bounds on the information rates
of a multi-sample receiver in Sec. \ref{sec:lower-bound}.
In Sec. \ref{sec:num-sim}, we report the results of numerical simulations
and Sec. \ref{sec:conc} concludes the paper.

\section{Continuous-Time Model}
\label{sec:ct-model}
We use the following notation: 
$j=\sqrt{-1}$ , 
$^*$ denotes the complex conjugate, 
$\delta_D$ is the Dirac delta function, 
$\lceil \cdot \rceil$ is the ceiling operator.
We use $X^k$ to denote $(X_1,X_2,\ldots,X_k)$.
Suppose the transmit-waveform is $x(t)$ and 
the receiver observes 
\begin{align}
  r(t) = x(t) \ e^{j \theta(t)} + n(t)
  \label{eq:rx_waveform}
\end{align}
where $n(t)$ is a realization of a white circularly-symmetric complex Gaussian process $N(t)$ with
\begin{align}
&\mathbb{E}\left[ N(t) \right] = 0 \nonumber \\
&\mathbb{E}\left[ N(t_1) N^*(t_2) \right] = \sigma^2_N ~ \delta_D(t_2-t_1).
\label{eq:Nt_statistics}
\end{align}
The phase $\theta(t)$ is a realization of a Wiener process $\Theta(t)$:
\begin{align}
  \Theta(t) = \Theta(0) + \int_0^t W(\tau) d\tau
\label{eq:Thetat}
\end{align}
where $\Theta(0)$ is uniform  on $[-\pi,\pi)$ and 
$W(t)$ is a real Gaussian process with
\begin{align}
&\mathbb{E}\left[ W(t) \right] = 0 \nonumber \\
&\mathbb{E}\left[ W(t_1) W(t_2)\right] = 2\pi \beta ~ \delta_D(t_2-t_1) .
\label{eq:Wt_statistics}
\end{align}
The processes $N(t)$ and $\Theta(t)$ are independent of each other and independent of the input.
$N_0 = 2 \sigma^2_N$ is the single-sided power spectral density of the additive noise.
We define
$U(t) \equiv \exp(j \Theta(t))$.
The autocorrelation function of $U(t)$ is
\begin{align}
R_U(t_1,t_2) 
= \mathbb{E}\left[ U(t_1) U^*(t_2)\right] 
= \exp\left(- \pi \beta |t_2-t_1| \right)
\end{align}
and the power spectral density of $U(t)$ is
\begin{align}
S_U(f)
= \int_{-\infty }^{\infty} R_U(t,t+\tau) \ e^{-j 2\pi f \tau} d\tau
= \frac{\beta/2}{(\beta/2)^2+ f^2}
\end{align}
The spectrum is said to have a Lorentzian shape.
It is easy to show that
$\beta = f_{\text{FWHM}} = 2 f_{\text{HWHM}}$
where 
$f_{\text{FWHM}}$ is the full-width at half-maximum
and
$f_{\text{HWHM}}$ is the half-width at half-maximum.
Let $T$ be the transmission interval, then
the transmitted waveforms must satisfy the power constraint
\begin{align}
	\mathbb{E}\left[\frac{1}{T} \int_0^{T} |X(t)|^2 dt \right] \leq \mathcal{P}
	\label{eq:finitesupport_waveform_power_constraint}
\end{align}
where $X(t)$ is a random process whose realization is $x(t)$.

\section{Discrete-Time Model}
\label{sec:dt-model}
Let $(\xsymb{1},\xsymb{1},\ldots,\xsymb{\blk})$
be the codeword sent by the transmitter.
Suppose the transmitter uses a unit-energy pulse $g(t)$ whose time support is $[0,\Ts]$
where $\Ts$ is the symbol interval.
The waveform sent by the transmitter is
\begin{align}
 x(t) = \sum_{m=1}^{\blk}  \xsymb{m} \ g(t-(m-1) \Ts).
\label{eq:xt_modulated_rect}
\end{align}
Let $L$ be the number of samples per symbol ($L \geq 1$)
and define the sample interval as
\begin{align}
\Delta = \frac{\Ts}{L}.
\end{align}

The received waveform $r(t)$ is filtered using an integrator over a sample interval to give the output signal
\begin{align}
y(t) 
&= \int_{t - \Delta}^{t} r(\tau) \ d\tau.
\label{eq:yt}
\end{align}
The signal $y(t)$ is a realization of $Y(t)$ that
is sampled at $t = k \Delta$, 
$k = 1, \ldots, \nsamp = \nsymb L$,
to yield the discrete-time model:
\begin{align}
Y_k = \Xsymb{\lceil k/L \rceil} \Delta \ e^{j \Theta_k} \ F_k + N_k
\label{eq:Yk}
\end{align}
where
$Y_k \equiv Y(k \Delta)$, 
$\Theta_k \equiv \Theta( (k-1) \Delta )$,
\begin{align}
F_k \equiv \frac{1}{\Delta} \int_{(k-1) \Delta}^{k \Delta} 
g\left(\tau-\left(\left\lceil\frac{k}{L}\right\rceil -1\right) \Ts\right)
e^{j(\Theta(\tau)-\Theta_k)} \ d\tau
\label{eq:Fk_def}
\end{align}
and
\begin{align}
N_k 
&\equiv \int_{(k-1) \Delta}^{k \Delta} N(\tau) \ d\tau.
\label{eq:Nk_def}
\end{align}
The process $\{N_k\}$ is an i.i.d. circularly-symmetric complex Gaussian process with mean $0$ and 
$\mathbb{E}[ |N_k|^2 ] = \sigma^2_N \Delta$
while the process $\{\Theta_k\}$ is the discrete-time Wiener process:
\begin{align}
  \Theta_k = \Theta_{k-1} + W_k ~ \mod 2 \pi
\end{align}
for $k = 2, \ldots, n$,
where
$\Theta_1$ is uniform on $[-\pi,\pi)$ and
$\{W_k\}$ is an i.i.d. real Gaussian process with mean $0$ and $\mathbb{E}[ |W_k|^2 ] = 2\pi \beta \Delta$, i.e., 
the probability distribution function (pdf) of $W_k$ is $p_{W_k}(w) = G(w;0,\sigma^2_W)$ where
\begin{align}
 G(w;\mu,\sigma^2) = \frac{1}{\sqrt{2\pi \sigma^2}} \exp\left( - \frac{(w-\mu)^2}{2 \sigma^2} \right)
\end{align}
and $\sigma^2_W = 2 \pi \beta \Delta$.
The random variable $(W_k \mod 2\pi)$  is a \emph{wrapped Gaussian} and its pdf is $p_{W}(w;\sigma^2_W)$
where
\begin{align}
 p_{W}(w;\sigma^2) = \sum_{i=-\infty}^{\infty} G(w-2 i \pi;0,\sigma^2).
\end{align}
Moreover, $\{F_k\}$ and $\{W_k\}$ are independent of $\{N_k\}$ but not independent of each other.
Finally, equations (\ref{eq:finitesupport_waveform_power_constraint}) and (\ref{eq:xt_modulated_rect})
imply the power constraint 
\begin{align}
	\frac{1}{\blk} \sum_{m=1}^{\blk} \mathbb{E}[|\Xsymb{m}|^2] \leq P = \mathcal{P} \Ts.
	\label{eq:dt_power_constraint}
\end{align}

It is convenient to define $X_k$ as
\begin{align}
X_k \equiv X(k \Delta) = \Xsymb{\lceil k/L \rceil} ~ g\left( (k \text{ mod } L) \Delta\right).
\label{eq:Xk_def}
\end{align}
It follows that $I(\Xsymbol^\nsymb;Y^n) = I(X^n;Y^n)$.
We define the information rate 
\begin{align}
  I(X;Y) 
  = \lim_{\nsymb \rightarrow \infty} \frac{1}{\nsymb} I(X^n;Y^n).
\label{eq:IXY}
\end{align}
One difficuly in evaluating (\ref{eq:IXY}) is that the joint distribution of $\{F_k\}$ and $\{W_k\}$ is not available in closed form.
Even the distribution of $F_k$ is not available in closed form
(there is an approximation for small linewidth, see (16) in \cite{Foschini1988Comm}).
However, we can numerically compute tight lower bounds on $I(X;Y)$ by using the auxiliary-channel technique described next.

\section{Lower Bound}
\label{sec:lower-bound}
The Auxiliary-Channel Lower Bound Theorem in \cite[Sec. VI]{Arnold2006} states that
for two random variables $X$ and $Y$, we have
\begin{align}
I(X;Y)
&\geq \underline{I}(X;Y) 
= \mathbb{E}\left[\log\left( \frac{q_{Y|X}(Y|X)}{q_Y(Y)} \right)\right]
\label{eq:I_X_Y_LB_AuxCh}
\end{align}
where $q_{Y|X}(\cdot|\cdot)$ is an arbitrary auxiliary channel and
\begin{align}
 q_Y(y) = \sum_{\tilde{x}} p_{X}(\tilde{x}) q_{Y|X}(y|\tilde{x})
\end{align}
where $p_{X}$ is the \emph{true} distribution of $X$. 
The distribution $q_Y(\cdot)$ is thus the output distribution obtained by connecting the true input source to the auxiliary channel.
Using this theorem, we can compute a lower bound on $I(X;Y)$ by using the following algorithm \cite{Arnold2006}:
\begin{enumerate}
\item Sample a long sequence $(x^n, y^n)$ according to the \emph{true} joint distribution of $X^n$ and $Y^n$. 
\item Compute $q_{Y^n|X^n}(y^n|x^n)$
and
\begin{align}
 q_{Y^n}(y^n) = \sum_{\tilde{x}^n} p_{X^n}(\tilde{x}^n) q_{Y^n|X^n}(y^n|\tilde{x}^n)
\end{align}
where $p_{X^n}$ is the true distribution of $X^n$.
\item Estimate $\underline{I}(X;Y)$ using
\begin{align}
  \underline{I}(X;Y) \approx 
  \frac{1}{\nsymb}\log\left( \frac{q_{Y^n|X^n}(y^n|x^n)}{q_{Y^n}(y^n)} \right)
\end{align}
\end{enumerate}

\paragraph*{Auxiliary Channel I}
Consider the auxiliary channel
\begin{align}
\Psi_k 
= X_k \Delta \ e^{j \Theta_k} + N_k
\label{eq:Psik_def}
\end{align}
where $\{\Theta_k\}$ and $\{N_k\}$ are defined in Sec. \ref{sec:dt-model}
and
$X_k$ is defined by (\ref{eq:Xk_def}).
The channel $\Psi$ is the same as $Y$ in (\ref{eq:Yk}) \emph{except} that $F_k$ is replaced with $g\left( (k \mod L) \Delta\right)$.
The channel is described by
the conditional distribution $p_{\Psi^n|X^n}$
\begin{align}
  p_{\Psi^n|X^n}(y^n|x^n) 
  &=\int_{\theta^n} p_{\Theta^n,\Psi^n|X^n}(\theta^n,y^n|x^n) \ d\theta^n
\end{align}
where
\begin{align}
  &p_{\Theta^n,\Psi^n|X^n}(\theta^n,y^n|x^n) \nonumber \\ \qquad
  &= \prod_{k=1}^{n} p_{\Theta_k|\Theta_{k-1}}(\theta_k|\theta_{k-1}) \ 
     p_{\Psi|X,\Theta}(y_k|x_k,\theta_k)
  \label{eq:p_Theta_Psi|X}
\end{align}
with
\begin{align}
p_{\Theta_k|\Theta_{k-1}}(\theta|\tilde{\theta}) = 
\left\{
\begin{array}{ll}
p_{W}(\theta-\tilde{\theta}; \sigma^2_W), 	& k \geq 2 \\
1/(2\pi), 					& k = 1 
\end{array}
\right.
\end{align}
and
\begin{align}
p_{\Psi|X,\Theta}(y|x,\theta)
= \frac{1}{\pi \sigma^2_N \Delta} \exp\left(- \frac{\left|y - x ~ e^{j \theta} \right|^2}{\sigma^2_N \Delta} \right).
\label{eq:p_Y|XTheta}
\end{align}
The channel $p_{\Psi^n|X^n}$ has continuous states $\theta^n$,
which makes step 2 of the algorithm computationally infeasible.

\paragraph*{Auxiliary Channel II}
We use the following auxiliary channel for the numerical simulations:
\begin{align}
\Upsilon_k 
= X_k \Delta \ e^{j S_k} + N_k
\label{eq:Upsilonk_def}
\end{align}
which has the conditional probability
\begin{align}
  p_{\Upsilon^n|X^n}(y^n|x^n) 
  &=\sum_{s^n \in \mathcal{S}^n} p_{S^n,\Upsilon^n|X^n}(s^n,y^n|x^n)
\end{align}
where $\mathcal{S}$ is a \emph{finite} set and
\begin{align}
  &p_{S^n,\Upsilon^n|X^n}(s^n,y^n|x^n) \nonumber \\ \qquad
  &= \prod_{k=1}^{n} p_{S_k|S_{k-1}}(s_k|s_{k-1}) \ 
     p_{\Psi|X,\Theta}(y_k|x_k,s_k)
  \label{eq:p_S_Upsilon|X}
\end{align}
where
\begin{align}
p_{S_k|S_{k-1}}(s|\tilde{s}) = 
\left\{
\begin{array}{ll}
Q(s|\tilde{s}), 			& k \geq 2 \\
1/|\mathcal{S}|,			& k = 1. 
\end{array}
\right.
\end{align}

Next, we describe our choice of $\mathcal{S}$ and $Q(\cdot|\cdot)$.
We  partition $[-\pi,\pi)$ into $S$ intervals with equal lengths
and pick the mid points of these intervals to be the elements of $\mathcal{S}$,
i.e., we have
\begin{align}
\mathcal{S} = \left\{ \hat{s}_i : i = 1,\ldots,S \right\}
\text{ where }
\hat{s}_i = i \frac{2 \pi}{S} - \frac{\pi}{S} - \pi
.
\end{align}
The state transition probability $Q(\cdot|\cdot)$ is chosen similar to \cite{BarlettaLB} and \cite{BarlettaUB}:
\begin{align}
Q(s|\tilde{s}) = 
\frac{2\pi}{S}
{\int_{(\phi,\tilde{\phi}) \in \mathcal{R}(s) \times \mathcal{R}(\tilde{s})} 
p_{W}(\phi-\tilde{\phi};\sigma^2_W) \ d\phi d\tilde{\phi}}
\end{align}
where $\mathcal{R}(s) = [ s- {\pi}/{S} , s + {\pi}/{S} )$, i.e.,
$\mathcal{R}(s)$ is the interval whose midpoint is $s$.
The larger $S$ and $L$ are, the better the auxiliary channel (\ref{eq:Upsilonk_def}) approximates the actual channel (\ref{eq:Yk}).
We remark that even for small $S$ and $L$, the auxiliary channel gives a \emph{valid} lower bound on $I(X;Y)$.

\subsection{Computing The Conditional Probability}
\begin{figure}
\resizebox{\columnwidth}{!}{\input{bayesian_net_general}}
\caption{Bayesian network for ${X^n,S^n,\Upsilon^n}$ for $n=9$.}
\label{fig:bayesian_net_general}
\end{figure}
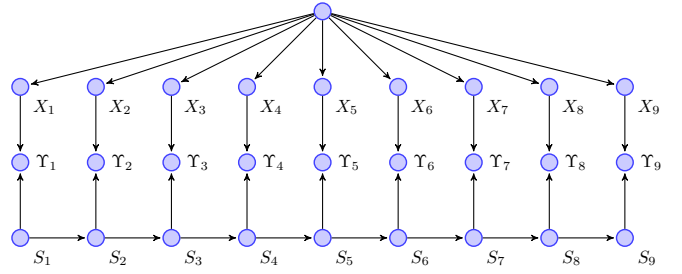

Suppose the input $X^n$ has the distribution $p_{X^n}$.
A Bayesian network for $X^n,S^n,\Upsilon^n$ is shown in Fig. \ref{fig:bayesian_net_general}.
The probability $p_{\Upsilon^n|X^n}(y^n|x^n)$ can be computed using
\begin{align}
 p_{\Upsilon^n|X^n}(y^n|x^n) 
 = \sum_{s \in \mathcal{S}} \rho_n(s)
\end{align}
where we recursively compute 
\begin{align}
  \rho_{k}(s) 
  & \equiv p_{S_k,\Upsilon^k|X^n}(s,y^k|x^n) \label{eq:rho_def} \\
  &\stackrel{(a)}{=}  \sum_{\tilde{s} \in \mathcal{S}} p_{S_{k-1},S_k,\Upsilon^k|X^n}(\tilde{s},s,y^k|x^n) \nonumber \\
  &\stackrel{(b)}{=}  \sum_{\tilde{s} \in \mathcal{S}} 
      \rho_{k-1}(\tilde{s}) \ 
      p_{S_k,\Upsilon_k|S_{k-1},\Upsilon^{k-1},X^n}(s,y_k|\tilde{s},y^{k-1},x^n) \nonumber \\
  &=  \sum_{\tilde{s} \in \mathcal{S}} 
      \rho_{k-1}(\tilde{s}) \ 
      Q(s|\tilde{s}) \ 
      p_{\Psi|X,\Theta}(y_k|x_k,s)
  \label{eq:rho_recursive_general}
\end{align}
with the initial value
$\rho_{0}(s) = 1/|\mathcal{S}|$.
Step $(a)$ is a marginalization, 
$(b)$ follows from Bayes' rule and the definition of $\rho_k$ in (\ref{eq:rho_def}), 
while (\ref{eq:rho_recursive_general}) follows from the structure of Fig. \ref{fig:bayesian_net_general}.
We remark that (\ref{eq:rho_recursive_general}) is the same as with independent $X_1,\ldots,X_n$,
e.g., see equation (9) in \cite[Sec. IV]{MTR2013}.

\subsection{Computing The Marginal Probability}
\begin{figure}
\resizebox{\columnwidth}{!}{\input{bayesian_net_oversamp}}
\caption{Bayesian network for ${X^n,S^n,\Upsilon^n}$ for $n=9$ and $L=3$.}
\label{fig:bayesian_net_oversamp}
\end{figure}
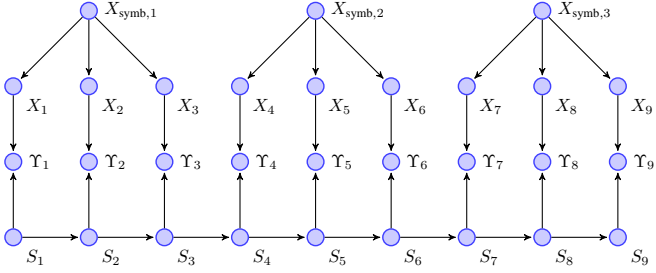

Define $\Y_m \equiv (Y_{(m-1) L + 1},Y_{(m-1) L + 2},\ldots,Y_{(m-1) L + L})$
and $\X_m \equiv (X_{(m-1) L + 1},X_{(m-1) L + 2},\ldots,X_{(m-1) L + L})$.
Suppose the input \emph{symbols} are i.i.d. and
$\Xsymb{m} \in \mathcal{X}$ where $\mathcal{X}$ is a finite set.
Therefore, $p_{X^n}$ has the form
\begin{align}
  p_{X^n}(x^n)
  &= \prod_{m=1}^{\blk} p_{\X}(\x_m).
\end{align}
A Bayesian network for $X^n,S^n,\Upsilon^n$ is shown in Fig. \ref{fig:bayesian_net_oversamp}.
The probability $p_{\Upsilon^n}(y^n)$ can be computed using
\begin{align}
  p_{\Upsilon^n}(y^n) 
  = \sum_{s \in \mathcal{S}} \psi_{\blk}(s) 
\end{align}
where $\psi_{m}(s) \equiv p_{S_{m L},\Y^m}(s,\y^m) $
which can be computed using the recursion:
\begin{align}
  &\psi_{m}(s) \label{eq:psi_recursive} \\
  &=  \sum_{\tilde{\x} \in \mathcal{X}_L} p_{\X}(\tilde{\x}) \
      \sum_{\tilde{s} \in \mathcal{S}} \psi_{m-1}(\tilde{s}) \
      p_{S_{m L},\Y_m|S_{(m-1)L},\X_m}(s,\y_m|\tilde{s},\tilde{\x})
  \nonumber
\end{align}
with the initial value 
$\psi_{0}(s) = 1/|\mathcal{S}|$.
The set $\mathcal{X}_L$ is
\begin{align}
 \mathcal{X}_L = \{ x \cdot (g(\Delta),g(2\Delta),\ldots,g(L\Delta) ): x \in \mathcal{X} \}.
\end{align}
We remark that $|\mathcal{X}_L| = |\mathcal{X}|$ and not $|\mathcal{X}|^L$.
Next, we define
\begin{align}
  \chi_{m,L}(s,\tilde{s},\tilde{\x}) \equiv p_{S_{m L},\Y_m|S_{(m-1)L},\X_m}(s,\y_m|\tilde{s},\tilde{\x}) 
  \label{eq:chi_def}
\end{align}
for $s,\tilde{s} \in \mathcal{S}$ and $\tilde{\x} \in \mathcal{X}_L$.
Computing $\chi_{m,L}(s,\tilde{s},\tilde{\x})$ is similar
to computing $\rho_n$ (see (\ref{eq:rho_recursive_general})).
Intuitively, this is because a block of $L$ samples in Fig. \ref{fig:bayesian_net_oversamp} has a structure similar to Fig. \ref{fig:bayesian_net_general}.
More precisely, 
$\chi_{m,L}(s,\tilde{s},\tilde{\x})$ can be computed recursively by using
\begin{align}
  &\chi_{m,\ell}(s,\tilde{s},\tilde{\x}) \label{eq:chi_recursive} \\
  &= \sum_{ \varsigma \in \mathcal{S} } 
     \chi_{m,\ell-1}(\varsigma,\tilde{s},\tilde{\x}) \ 
     Q\left( s| \varsigma \right) \
     p_{\Psi|X,\Theta} \left( y_{(m-1)L+\ell}|\tilde{x}_{\ell}, s\right)
     \nonumber
\end{align}
with the initial value
\begin{align}
  \chi_{m,0}(s,\tilde{s},\tilde{\x}) = \left\{
  \begin{array}{ll}
   1, & s=\tilde{s} \\
   0, & \text{otherwise}.
  \end{array}
  \right.
  \label{eq:chi_init}
\end{align}

Therefore, computing $p_{\Upsilon^n}(y^n)$ involves two levels of recursion:
1) recursion over the symbols as described by (\ref{eq:psi_recursive}) and 
2) recursion over the samples within a symbol as described by (\ref{eq:chi_recursive}).

\section{Numerical Simulations}
\label{sec:num-sim}
We use two pulses with a symbol-interval time support:
\begin{itemize}
\item 
A unit-energy square pulse
\begin{align}
g_1(t) = 
\frac{1}{\sqrt{\Ts}} \text{rect}\left(\frac{t}{\Ts}\right)
\end{align}
where 
\begin{align}
 \text{rect}(t) \equiv \left\{ 
  \begin{array}{ll}
  1,	& |t| \leq 1/2 , \\
  0,	& \text{otherwise}.
  \end{array}
 \right.
\label{eq:rect_def}
\end{align}

\item 
A unit-energy cosine-squared pulse
\begin{align}
g_2(t) = 
\frac{1}{\sqrt{\Ts/2}} \cos^2\left(\frac{\pi t}{\Ts}\right) \text{rect}\left(\frac{t}{\Ts}\right).
\end{align}
\end{itemize}

The first step of the algorithm is to sample a long sequence according to the true joint distribution of $X^n$ and $Y^n$. 
To generate samples according to the original channel (\ref{eq:Yk}),
we must accurately represent digitally the continuous-time waveform (\ref{eq:rx_waveform}).
We use a simulation oversampling rate $L_{\text{sim}}$ = 1024 samples/symbol.
After the filter (\ref{eq:yt}), the receiver has $L$ samples/symbol distributed according to (\ref{eq:Yk}).
Next, to choose a proper sequence length, we follow the approach suggested in \cite{Arnold2006}:
for a candidate length, run the algorithm about 10 times (each with a new random seed)
and check whether all estimates of the information rate agree up to the desired accuracy.
We used $\nsymb = 10^4$ unless otherwise stated.
We define the signal-to-noise ratio as $\SNR \equiv {P}/{\sigma^2_N \Ts} = {\mathcal{P}}/{\sigma^2_N}$.

For efficient implementation of (\ref{eq:rho_recursive_general}), 
$p_{\Psi|X,\Theta}(\cdot|\cdot,\cdot)$  
can be factored out of the summation to yield:
\begin{align}
  \rho_{k}(s) 
  &=  p_{\Psi|X,\Theta}(y_k|x_k,s)
      \overbrace{\sum_{\tilde{s} \in \mathcal{S}} \rho_{k-1}(\tilde{s}) \ Q(s|\tilde{s})}^{\rho^\prime_{k}(s)}
  \label{eq:rho_recursive_general_implement}
\end{align}
Moreover, 
since $Q(\cdot|\cdot)$ can be represented by a circulant matrix due to symmetry,
$\rho^\prime_k(\cdot)$ can be computed efficiently using the Fast Fourier Transform (FFT).
Similarly, the computation of (\ref{eq:chi_recursive}) can be done efficiently  
by factoring out $p_{\Psi|X,\Theta}(\cdot|\cdot,\cdot)$ and
by using the FFT.

\subsection{Excessively Large Linewidth}

\begin{figure}[ht]
\centering
\includegraphics[width=0.5\textwidth]{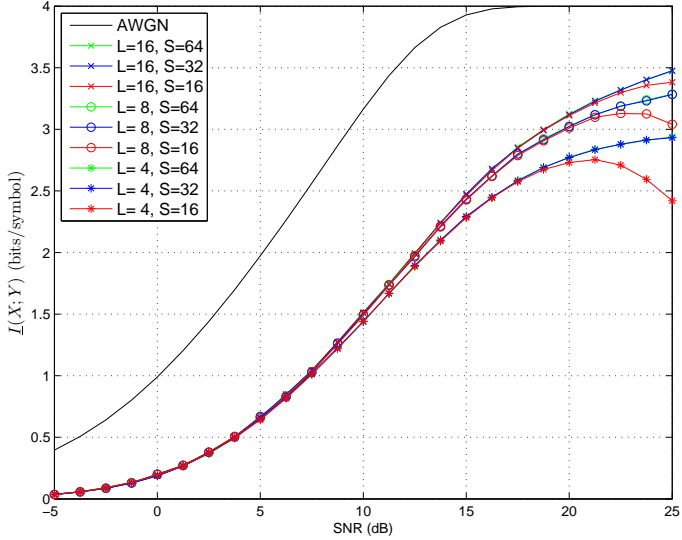}
\caption{Lower bounds on rates for 16-QAM, square transmit-pulse and multi-sample receiver at $f_{\text{HWHM}} \Ts = 0.125$.}
\label{fig:16qam_fhwhm125_sqrtx_optrx_osr}
\end{figure}
Suppose $f_{\text{HWHM}} \Ts = 0.125$ and the input symbols are independently and uniformly distributed (i.u.d.) 16-QAM.
Fig. \ref{fig:16qam_fhwhm125_sqrtx_optrx_osr} shows an estimate of $\underline{I}(X;Y)$ 
for a square transmit-pulse, 
i.e., $g(t) = g_1(t-\Ts/2)$
and an $L$-sample receiver  
with $L=4,8,16$ and $S=16,32,64$.
The curves with $S=64$ are indistinguishable from the curves with $S=32$ over the entire SNR range for all values of $L$,
and hence $S=32$ is adequate up to 25 dB.
Even $S=16$ is adequate up to 20 dB.
The important trend in Fig. \ref{fig:16qam_fhwhm125_sqrtx_optrx_osr}
is that higher oversampling rate $L$ is needed at high $\SNR$
to extract all the information from the received signal.
For example,
$L=4$ suffices up to $\SNR$ $\sim$ 10 dB,
$L=8$ suffices up to $\SNR$ $\sim$ 15 dB but
$L \geq 16$ is needed beyond that.
It was pointed out in \cite{Arnold2006} that the lower bounds can be interpeted as 
the information rates achieved by mismatched decoding.
For example, $\underline{I}(X;Y)$ for $L=8$ and $S \geq 32$ in Fig. \ref{fig:16qam_fhwhm125_sqrtx_optrx_osr} is essentially
the information rate achieved by a multi-sample (8-sample) receiver
that uses maximum-likelihood decoding for the simplified channel (\ref{eq:Psik_def})
when it is operated in the original channel (\ref{eq:Yk}).

\begin{figure}[ht]
\centering
\includegraphics[width=0.5\textwidth]{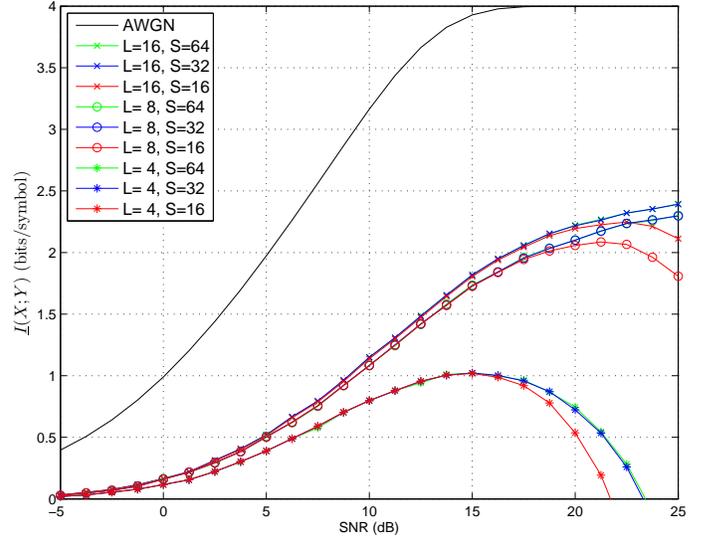}
\caption{Lower bounds on rates for 16-QAM, cosine-squared transmit-pulse and multi-sample receiver at $f_{\text{HWHM}} \Ts = 0.125$.}
\label{fig:16qam_fhwhm125_cos2tx_optrx_osr}
\end{figure}

Fig. \ref{fig:16qam_fhwhm125_cos2tx_optrx_osr} shows an estimate of $\underline{I}(X;Y)$ 
for a cosine-squared transmit-pulse, 
i.e., $g(t) = g_2(t-\Ts/2)$
and an $L$-sample receiver  
at $L=4,8,16$ and $S=16,32,64$.
We find that $S = 32$ suffices up to $\sim$ 25 dB.
We see in Fig. \ref{fig:16qam_fhwhm125_cos2tx_optrx_osr}
the same trend in Fig. \ref{fig:16qam_fhwhm125_sqrtx_optrx_osr}:
higher $L$ is needed at higher $\SNR$.
Comparing Fig. \ref{fig:16qam_fhwhm125_sqrtx_optrx_osr} with Fig. \ref{fig:16qam_fhwhm125_cos2tx_optrx_osr} indicates that 
the square pulse is better than the cosine-squared pulse for the same oversampling rate $L$.

\subsection{Large Linewidth}

\begin{figure}[ht]
\centering
\includegraphics[width=0.5\textwidth]{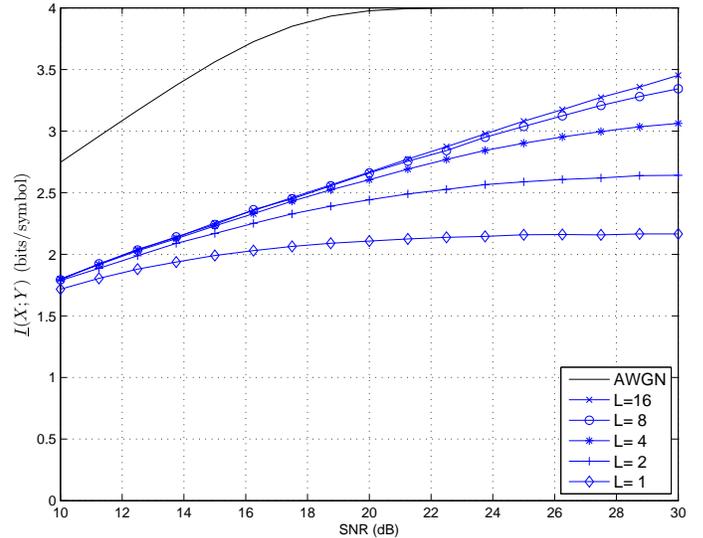}
\caption{Lower bounds on rates for 16-PSK, square transmit-pulse and multi-sample receiver at $f_{\text{HWHM}} \Ts = 0.0125$.}
\label{fig:16psk_fhwhm12_sqrtx_optrx_osr}
\end{figure}


As the linewidth decreases, the benefit of oversampling at the receiver becomes apparant only at higher $\SNR$.
For example, for $f_{\text{HWHM}} \Ts = 0.0125$ and i.u.d. 16-PSK input,
Fig. \ref{fig:16psk_fhwhm12_sqrtx_optrx_osr} shows an estimate of $\underline{I}(X;Y)$ 
for a square transmit-pulse 
and an $L$-sample receiver  
at $L=1,2,4,8,16$ and $S=64$.
We see that
$L=4$ suffices up to $\SNR$ $\sim$ 19 dB,
$L=8$ suffices up to $\SNR$ $\sim$ 24 dB and
only beyond that $L \geq 16$ is necessary.

We conclude from Fig. \ref{fig:16qam_fhwhm125_sqrtx_optrx_osr}--\ref{fig:16psk_fhwhm12_sqrtx_optrx_osr} that
the required $L$ depends on
1) the linewidth $f_{\text{FWHM}}$ of the phase noise;
2) the pulse $g(t)$; and
3) the $\SNR$.

\subsection{Comparison With Other Models}
We compare the discrete-time model of the multi-sample receiver
with other discrete-time models.
The simulation parameters for our model (GK) are $\nsymb = 10^4$, $L = 16$ (with $L_{\text{sim}} = 1024$) and 
$S=64$ for 16-QAM ($S=128$ was too computationally intensive) and $S=128$ for QPSK.

\begin{figure}[ht]
\centering
\includegraphics[width=0.5\textwidth]{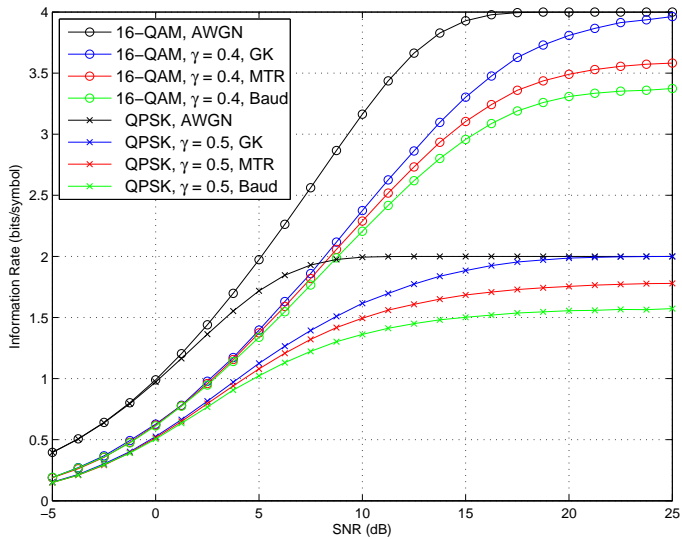}
\caption{Comparison of information rates for different models.}
\label{fig:raheli_comparison}
\end{figure}

In Fig. \ref{fig:raheli_comparison}, we show curves
for the Baud-rate model used in 
\cite{Barbieri2011} and \cite{Dauwels2008}--\cite{Barletta2012KalmanLB}.
The model is (\ref{eq:dt-pn-ch}) where the phase noise is a Wiener process whose noise increments have variance $\gamma^2$.
We set $\gamma^2 = 2 \pi \beta \Ts$.
The simulation parameters for the Baud-rate model are  $\nsymb = 10^5$ and $S=128$.

We also show curves for the Martal\`{o}-Tripodi-Raheli (MTR) model \cite{MTR2013} in Fig. \ref{fig:raheli_comparison}.
For the sake of comparison, we adapt the model in \cite{MTR2013} 
from a square-root raised-cosine pulse to a square pulse 
and write the ``matched'' filter output $\{V_m\}$ as
\begin{align}
 V_m = \sum_{\ell = 1}^L \Psi_{(m-1)L+1}
\end{align}
where $m=1,\ldots,\nsymb$ and $\Psi_k$ is defined in (\ref{eq:Psik_def}).
The auxiliary channel is 
\begin{align}
 Y_m = \Xsymb{m} ~ e^{j \Theta_m} + Z_m,
 \qquad m \geq 1
\end{align}
where 
the process $\{Z_m\}$ is an i.i.d. circularly-symmetric complex Gaussian process with mean $0$ and 
$\mathbb{E}[ |Z_m|^2 ] = \sigma^2_N \Ts$
while the process $\{\Theta_m\}$ is a first-order Markov process (not a Wiener process)
with a time-invariant transition probability, 
i.e., for $k \geq 2$ and all $\theta_k,\theta_{k-1} \in [-\pi,\pi)$, we have
$p_{\Theta_k|\Theta_{k-1}}(\theta_k|\theta_{k-1}) = p_{\Theta_2|\Theta_1}(\theta_k|\theta_{k-1})$.
Furthermore, the phase space is quantized to a finite number $S$ of states
and the transition probabilities are estimated by means of simulation.
The probabilities are then used to compute a lower bound on the information rate.
The simulation parameters for the MTR model are  $\nsymb = 10^5$, $L=16$ and $S=128$.

We see that the Baud-rate and MTR models saturate at a rate well below
the rate achieved by the multi-sample receiver.
Moreover, the multi-sample receiver achieves 
the full 4 bits/symbol and 2 bits/symbol of 16-QAM and QPSK, respectively, at high SNR.

\section{Conclusion}
\label{sec:conc}
We studied a waveform channel impaired by Wiener phase noise and AWGN
by evaluating via numerical simulations tight lower bounds on the information rates achieved by a multi-sample receiver.
We found that the required oversampling rate depends on
the linewidth of the phase noise, the shape of the transmit-pulse and the signal-to-noise ratio.
The results demonstrate that multi-sample receivers increase the information rate
for both strong and weak phase noise at high SNR.
We compared our results with the results obtained by using other discrete-time models.

\section*{Acknowledgment}
H. Ghozlan was supported by a USC Annenberg Fellowship and NSF Grant CCF-09-05235.
G. Kramer was supported by an Alexander von Humboldt Professorship endowed by
the German Federal Ministry of Education and Research.
The use of the FFT was suggested to us by L. Barletta.

\bibliographystyle{IEEEtran}
\bibliography{ref9}

\end{document}

%% file: bayesian_net_general.tex
\begin{tikzpicture}[->,>=stealth',shorten >=1pt,auto,node distance=1.5cm,semithick]
  \tikzstyle{var}=[circle,thick,draw=blue!75,fill=blue!20,minimum size=3pt]

  \node[var]  (S1) [		label=-45:$S_1$] {};
  \node[var]  (S2) [right of=S1,label=-45:$S_2$] {};
  \node[var]  (S3) [right of=S2,label=-45:$S_3$] {};
  \node[var]  (S4) [right of=S3,label=-45:$S_4$] {};
  \node[var]  (S5) [right of=S4,label=-45:$S_5$] {};
  \node[var]  (S6) [right of=S5,label=-45:$S_6$] {};
  \node[var]  (S7) [right of=S6,label=-45:$S_7$] {};
  \node[var]  (S8) [right of=S7,label=-45:$S_8$] {};
  \node[var]  (S9) [right of=S8,label=-45:$S_9$] {};

  \node[var]  (Y1) [above of=S1,label=right:$\Upsilon_1$] {};
  \node[var]  (Y2) [right of=Y1,label=right:$\Upsilon_2$] {};
  \node[var]  (Y3) [right of=Y2,label=right:$\Upsilon_3$] {};
  \node[var]  (Y4) [right of=Y3,label=right:$\Upsilon_4$] {};
  \node[var]  (Y5) [right of=Y4,label=right:$\Upsilon_5$] {};
  \node[var]  (Y6) [right of=Y5,label=right:$\Upsilon_6$] {};
  \node[var]  (Y7) [right of=Y6,label=right:$\Upsilon_7$] {};
  \node[var]  (Y8) [right of=Y7,label=right:$\Upsilon_8$] {};
  \node[var]  (Y9) [right of=Y8,label=right:$\Upsilon_9$] {};

  \node[var]  (X1) [above of=Y1,label=-45:$X_1$] {};
  \node[var]  (X2) [right of=X1,label=-45:$X_2$] {};
  \node[var]  (X3) [right of=X2,label=-45:$X_3$] {};
  \node[var]  (X4) [right of=X3,label=-45:$X_4$] {};
  \node[var]  (X5) [right of=X4,label=-45:$X_5$] {};
  \node[var]  (X6) [right of=X5,label=-45:$X_6$] {};
  \node[var]  (X7) [right of=X6,label=-45:$X_7$] {};
  \node[var]  (X8) [right of=X7,label=-45:$X_8$] {};
  \node[var]  (X9) [right of=X8,label=-45:$X_9$] {};

  \node[var]  (W)  [above of=X5] {};

  \path 
  (S1) edge node {} (S2)
  (S2) edge node {} (S3)
  (S3) edge node {} (S4)
  (S4) edge node {} (S5)
  (S5) edge node {} (S6)
  (S6) edge node {} (S7)
  (S7) edge node {} (S8)
  (S8) edge node {} (S9)
  (X1) edge node {} (Y1)
  (X2) edge node {} (Y2)
  (X3) edge node {} (Y3)
  (X4) edge node {} (Y4)
  (X5) edge node {} (Y5)
  (X6) edge node {} (Y6)
  (X7) edge node {} (Y7)
  (X8) edge node {} (Y8)
  (X9) edge node {} (Y9)
  (S1) edge node {} (Y1)
  (S2) edge node {} (Y2)
  (S3) edge node {} (Y3)
  (S4) edge node {} (Y4)
  (S5) edge node {} (Y5)
  (S6) edge node {} (Y6)
  (S7) edge node {} (Y7)
  (S8) edge node {} (Y8)
  (S9) edge node {} (Y9)
  (W) edge node {} (X1)
  (W) edge node {} (X2)
  (W) edge node {} (X3)
  (W) edge node {} (X4)
  (W) edge node {} (X5)
  (W) edge node {} (X6)
  (W) edge node {} (X7)
  (W) edge node {} (X8)
  (W) edge node {} (X9)
;
\end{tikzpicture}

%% file: bayesian_net_oversamp.tex
\begin{tikzpicture}[->,>=stealth',shorten >=1pt,auto,node distance=1.5cm,semithick]
  \tikzstyle{var}=[circle,thick,draw=blue!75,fill=blue!20,minimum size=3pt]

  \node[var]  (S1) [		label=-45:$S_1$] {};
  \node[var]  (S2) [right of=S1,label=-45:$S_2$] {};
  \node[var]  (S3) [right of=S2,label=-45:$S_3$] {};
  \node[var]  (S4) [right of=S3,label=-45:$S_4$] {};
  \node[var]  (S5) [right of=S4,label=-45:$S_5$] {};
  \node[var]  (S6) [right of=S5,label=-45:$S_6$] {};
  \node[var]  (S7) [right of=S6,label=-45:$S_7$] {};
  \node[var]  (S8) [right of=S7,label=-45:$S_8$] {};
  \node[var]  (S9) [right of=S8,label=-45:$S_9$] {};

  \node[var]  (Y1) [above of=S1,label=right:$\Upsilon_1$] {};
  \node[var]  (Y2) [right of=Y1,label=right:$\Upsilon_2$] {};
  \node[var]  (Y3) [right of=Y2,label=right:$\Upsilon_3$] {};
  \node[var]  (Y4) [right of=Y3,label=right:$\Upsilon_4$] {};
  \node[var]  (Y5) [right of=Y4,label=right:$\Upsilon_5$] {};
  \node[var]  (Y6) [right of=Y5,label=right:$\Upsilon_6$] {};
  \node[var]  (Y7) [right of=Y6,label=right:$\Upsilon_7$] {};
  \node[var]  (Y8) [right of=Y7,label=right:$\Upsilon_8$] {};
  \node[var]  (Y9) [right of=Y8,label=right:$\Upsilon_9$] {};

  \node[var]  (X1) [above of=Y1,label=-45:$X_1$] {};
  \node[var]  (X2) [right of=X1,label=-45:$X_2$] {};
  \node[var]  (X3) [right of=X2,label=-45:$X_3$] {};
  \node[var]  (X4) [right of=X3,label=-45:$X_4$] {};
  \node[var]  (X5) [right of=X4,label=-45:$X_5$] {};
  \node[var]  (X6) [right of=X5,label=-45:$X_6$] {};
  \node[var]  (X7) [right of=X6,label=-45:$X_7$] {};
  \node[var]  (X8) [right of=X7,label=-45:$X_8$] {};
  \node[var]  (X9) [right of=X8,label=-45:$X_9$] {};

  \node[var]  (M1) [above of=X2,label=right:$X_{\text{symb},1}$] {};
  \node[var]  (M2) [above of=X5,label=right:$X_{\text{symb},2}$] {};
  \node[var]  (M3) [above of=X8,label=right:$X_{\text{symb},3}$] {};

  \path 
  (S1) edge node {} (S2)
  (S2) edge node {} (S3)
  (S3) edge node {} (S4)
  (S4) edge node {} (S5)
  (S5) edge node {} (S6)
  (S6) edge node {} (S7)
  (S7) edge node {} (S8)
  (S8) edge node {} (S9)
  (X1) edge node {} (Y1)
  (X2) edge node {} (Y2)
  (X3) edge node {} (Y3)
  (X4) edge node {} (Y4)
  (X5) edge node {} (Y5)
  (X6) edge node {} (Y6)
  (X7) edge node {} (Y7)
  (X8) edge node {} (Y8)
  (X9) edge node {} (Y9)
  (S1) edge node {} (Y1)
  (S2) edge node {} (Y2)
  (S3) edge node {} (Y3)
  (S4) edge node {} (Y4)
  (S5) edge node {} (Y5)
  (S6) edge node {} (Y6)
  (S7) edge node {} (Y7)
  (S8) edge node {} (Y8)
  (S9) edge node {} (Y9)
  (M1) edge node {} (X1)
  (M1) edge node {} (X2)
  (M1) edge node {} (X3)
  (M2) edge node {} (X4)
  (M2) edge node {} (X5)
  (M2) edge node {} (X6)
  (M3) edge node {} (X7)
  (M3) edge node {} (X8)
  (M3) edge node {} (X9)
;
\end{tikzpicture}